\documentclass[preprint,12pt]{elsarticle}

\usepackage{amsmath,amssymb,amsfonts}
\usepackage{braket}
\usepackage{algorithmic}
\usepackage{graphicx}
\usepackage{textcomp}
\usepackage{hyperref}
\usepackage{xcolor}
\usepackage{caption}
\usepackage{float}
\usepackage{subcaption}

\begin{document}

\begin{frontmatter}



\title{Fisher Information Measures under Lattice Combined Paul Trap}
\author[may1,thai]{Precious Ogbonda Amadi}
\author[thai]{Paphon Pewkhom}
\author[thai]{Pruet Kalasuwan}
	\ead{pruet.k@psu.ac.th}
\author[may2,may3]{Norshamsuri Ali}
\author[may1,may3]{Syed Alwee Aljunid}
\author[may1,may3]{Rosdisham Endut}

\affiliation[may1]{
		Faculty of Intelligent Computing, Universiti Malaysia Perlis,02600, Arau, Perlis, Malaysia}
\affiliation[thai]{Division of Physical Science, Faculty of Science,Prince of Songkhla University, Hat Yai, 90110, Songkhla, Thailand}

\affiliation[may2]{Faculty of Electronic Engineering & Technology, Universiti Malaysia Perlis,02600, Arau, Perlis, Malaysia} 

\affiliation[may3]{Centre of Excellence Advanced Communication Engineering (ACE), Universiti Malaysia Perlis,02600, Arau, Perlis, Malaysia} 

\cortext[1]{Corresponding author}

\begin{abstract}
We examine how the informational properties of a confined single ion response in a Paul trap modified by optical-lattice. We focus on the ground and first excited motional states and show that Fisher information, Shannon entropy, and Fisher-Shannon complexity track the effective frequency $\omega_{\mathrm{eff}}=\omega\,\sqrt{1-\kappa}$ of the potential. We show that the Fisher information and Shannon entropy reflect an effective frequency-driven redistribution of information between conjugate spaces. Our result show that the Fisher–Shannon complexity measure remains invariant under effective frequency control. The invariance demonstrates that optical modulation of $\kappa$ rescales localization, without altering the harmonic structure of the motional states. These results establish a controlled information-theoretic baseline for lattice-assisted Paul traps. Beyond the harmonic limit, retaining the quartic lattice correction introduces non-Gaussian wavefunction features through state-dependent mixing of higher eigenstates, which breaks the mutual compensation between Fisher information and Shannon entropy that sustains the invariant. The departure of $P^\prime$ from its harmonic reference value intensifies with $\kappa$ and is stronger for the 
excited state, which confirms that the Fisher-Shannon complexity invariance 
is a distinctive property of the small-oscillation harmonic regime.
\end{abstract}




\end{frontmatter}



\section{Introduction}
\label{sec1}

Trapped ions stand out as a leading platform in quantum optics and quantum information science. This is owed to their strong isolation from the environment, precision in control of quantum states, and long coherence times \cite{bruzewicz2019Trapped}. The choice of trapped ions as a preferred candidate for the development of quantum technology is hinged on their ability to confine single atomic ions in electromagnetic potentials. This confinement restricts the ion's motion to quantized vibrational modes, forming well-defined harmonic oscillator states \cite{leibfried2003quantum}. In strong confinement regimes, these vibrational levels are sharply spaced, and coherent laser-ion interactions provide full control over the internal and motional degrees of freedom for ground state cooling, state engineering, and quantum logic operations \cite{fogarty2016optomechanical,manovitz2022trapped,haffne2008quantum,monroe1997atomic}. 

Paul traps generate electromagnetic confinement that gives trapped-ion systems the characteristic of a harmonic oscillator potential. The electromagnetic confinement of charged particles arises from the rapidly oscillating radio-frequency quadrupole field. This field produces a stable, time-averaged pseudopotential whose secular motion of a single ion is accurately described by quantized vibrational levels \cite{landa2012modes,lindvall2022high}. The spacing of these levels is set by the strength of the confinement and determines the accessible motional dynamics \cite{huber2008employing, pan2020weak}. Working on a single trapped ion focuses on the pure quantum dynamics of confinement, in the absence of additional interactions and collective effects present in multi-ion chains. As such, the single atom trap ion system is well isolated from environmental disturbances, and both internal and motional states can be coherently manipulated and measured with high precision. Several research efforts demonstrate single-ion Paul traps as a controllable platform for quantum-optical and quantum-dynamical studies. Single trapped ion can be cooled near absolute zero and precisely controlled with laser light for accessibility of coherent control and measurement under well-controlled conditions  \cite{leibfried2003quantum}.  Experiments leveraging these capabilities have enabled high-fidelity quantum logic operations, sideband cooling to the motional ground state, and stable encoding of quantum information in long-lived internal states \cite{monroe1997atomic,holzl2023motional,shi25long}. Single atom trapped ion have been studied and experimentally realized as working media as the working substance for nano-scale and quantum heat engines \cite{abah2012single,ronagel2016single,david2018single}. Trapped ions serve as leading qubits in scalable quantum computers \cite{schwerdt2024scalable,harty014high}, and in precise sensing \cite{ivanov2016high}.  The success of these works rests on the fact that in a Paul trap, a single ion experiences a near-ideal harmonic potential with well-defined quantized vibrational levels, and the coupling between internal and motional states can be engineered precisely.

Recent research have introduced optical lattices as a versatile tool for shaping and controlling ion potentials with unprecedented spatial resolution and tunability. By superimposing a standing wave of laser light onto the radio-frequency pseudopotential of a Paul trap, researchers can create hybrid electrostatic-optical potentials that combine the long-term stability of electromagnetic confinement with the precise, wavelength-scale periodicity of optical potentials \cite{karpa2013supression,enderlein2012single}. This synthesis enables the realization of complex potential landscapes, such as double wells or periodic lattices with controllable depths and spacings \cite{david2018single,linnet2012pinning}. The resulting combined Paul trap lattice architecture allows for the engineering of inhomogeneous energy level shifts-a key ingredient for exploring novel quantum thermodynamic effects and friction dynamics at the atomic scale \cite{gangloff2015velocity, alexei2015tuning}. For instance, the introduction of a $\delta$-function-like barrier via an optical lattice can selectively modify quantum energy levels without affecting classical bulk properties. This modification leads to phenomena such as quantum-enhanced heat engine performance and the suppression of ion transport \cite{david2018single,karpa2013supression}. Furthermore, the interplay between the harmonic trap and the optical lattice provides a powerful platform for simulating nanofriction and studying stick-slip dynamics with single-atom resolution \cite{gangloff2015velocity, alexei2015tuning}.

Fisher information (FI) is a natural tool for analyzing how strong confinement shapes the structure of a quantum state \cite{wu2020fisher,mukherjee2017fisher}. The foundational works \cite{fisher1925theory,sears1980quantum} establish FI as a quantitative measure of spatial localization through its sensitivity to local gradients of the wavefunction. In quantum systems, FI captures how sharply the probability density responds to parameter variations. In our context, FI offers a precise way to track how optical effective frequency control redistributes localization properties of the motional state without altering its underlying harmonic structure. With significant progress in applying Fisher information to quantum systems, systematic analyses have focused primarily on central potentials, periodic systems, and canonical harmonic oscillators \cite{lima2023quantum,isonguyo2018quantum,ikot2020shannon,amadi2024complexity,inyang2025quantum,dong2018radial,li2024one,falaye2016fisher,santana-carrillo2023quantum, amadi2020shannon, onate2024fisher}. However, Fisher information analysis in a hybrid Paul-trap–lattice system under optically controlled frequency is yet to be reported. 

In this work, we investigate a single Paul-trapped ion subjected to an additional one-dimensional optical lattice in the strong-confinement regime. The combined potential introduces an effective harmonic form characterized by an effective frequency $\omega_{\mathrm{eff}}=\omega\sqrt{1-\kappa}$. Our objective is to establish a controlled information-theoretic baseline under optical effective frequency modulation. By expressing information theoretic measures such as Fisher information, Shannon entropy, and Fisher–Shannon complexity, we identify which information measures vary under effective frequency control through $\kappa$ and which remain invariant. It is important to clarify the scope of the present work. Our intention to reduce the effective harmonic confinement is not to introduce new eigen structure beyond that of the quantum harmonic oscillator. Rather, we show that the optical lattice parameter $\kappa$ acts as a pure scaling control on the effective harmonic confinement, $\omega_{\mathrm{eff}} = \omega\sqrt{1-\kappa}$, and that all three invariants therefore hold across the full range of optical tuning in the Paul-trap lattice. This establishes a concrete diagnostic baseline: any deviation from these invariants in a physical Paul-trap lattice experiment would signal non-harmonic lattice effects or higher-order corrections. Therefore, the invariants are not trivial in this context because $\kappa$ is an experimentally tunable optical parameter, not a fixed system property, and its role as a pure scaling handle has not been reported in the harmonic oscillator architecture.

The rest of the article is organized as follows: Section~\ref{model} discusses the harmonic Paul trap lattice model. We will analyze the frameworks of Fisher information in Section~\ref{fisher} and its complexity measure in Section~\ref{fishershannon} for our model. Section~\ref{result} will be for the discussion of our result, and conclusion Section~\ref{conclusion}.
\section{The Model}\label{model}
We consider a combined electrostatic potential consisting of the harmonic confinement of a Paul trap superimposed with a one-dimensional optical lattice~\cite{david2018single},
\begin{equation}\label{Eq01}
    V(x)=\frac{m \omega^2 x^2}{2}
   \, + \,\frac{m \omega^2 a^2\,\kappa}{4 \pi^2}
    \left[1+\cos\left(\frac{2 \pi x}{a}\right)\right],
\end{equation}
where \( a \) defines the spatial periodicity of the optical lattice. The parameter \( \omega \) denotes the secular vibrational frequency of the Paul trap. \( \kappa \) controls the relative contribution of the optical lattice to the overall potential. 

The interplay between \( \omega \) and \( \kappa \) captures the transition between weak and strong confinement regimes in the Paul trap–lattice system. In the weak-confinement regime, higher-order lattice terms dominates. This higher order leads to deviations from harmonic behavior and allows the ion wavefunction to extend over multiple lattice sites. These anharmonic effects influence both the energy spectrum and spatial coherence. On the other hand, under strong confinement, the ion remains tightly localized near a single lattice minimum, simplifying the theoretical description and enabling controlled analytical treatment relevant to  experiments~\cite{david2018single,gangloff2015velocity,koch2023quantum}.

\begin{figure}[!t]
    \centering
    \includegraphics[width=0.75\linewidth]{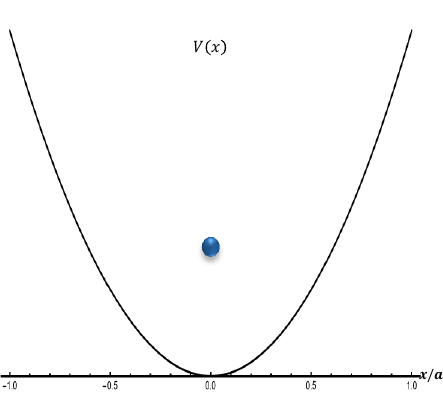}
    \caption{Harmonic Paul trap with combined optical lattice characterized by effective frequency $\omega_{\mathrm{eff}}=\omega\,\sqrt{1-\kappa}$ describes in  Eq.~\ref{eq02}}
    \label{fig:potmodel}
\end{figure}

We therefore consider the strong-confinement limit, where the ion motion is restricted to small displacements about the trap center, $x \ll a $, corresponding to a local minimum of the lattice potential. In this regime, the sinusoidal term in Eq.~\ref{Eq01} varies slowly and can be expanded in a Taylor series about $x=0$ as $\cos\!\left(2\pi x/a\right) = 1 - 1/2(2\pi x/a)^2
+ \mathcal{O}\!\left(x^4\right)$. We retain the terms up to second order in $x$, the potential reduces to:
\begin{equation}\label{eq02}
V(x)=\frac{m\omega_{\mathrm{eff}}^{2}}{2}\,x^{2}+\frac{ m\omega^{2}a^{2}\kappa}{ 2\pi^{2}},\qquad \omega_{\mathrm{eff}}=\omega\sqrt{1-\kappa}
\end{equation}
where $\kappa =0$, recovers the harmonic oscillator~\cite{sakurai2017modern}. The second term, the constant offset term, shifts the potential baseline without affecting the dynamics. It represents a zero-point energy offset arising from the lattice contribution. The second-order truncation isolates the leading restoring force and yields an effective harmonic confinement characterized by the effective frequency $\omega_{\mathrm{eff}}=\omega\sqrt{(1-\kappa)}$,
with $\kappa<1$, as shown in Fig.~\ref{fig:potmodel}. Varying $\kappa$ rescales the effective frequency continuously without modifying the trap settings. This allows effective frequency tuning to be implemented optically while preserving trap stability and micromotion conditions. In the rest of this work, $\kappa$  will be treated as an independent control optical parameter that determines $\omega_{\mathrm{eff}}$, while the underlying dynamical structure remains fixed.

In the strong-confinement regime, the combined Paul trap and optical lattice reduce to an effective harmonic potential characterized by $\omega_{\mathrm{eff}}=\omega\,\sqrt{1-\kappa}$. Varying \( \kappa \) rescales the effective frequency continuously without modifying the trap settings. This allows effective frequency tuning to be implemented optically while preserving trap stability and micromotion conditions. 


We consider the one-dimensional system described by the Hamiltonian
\begin{equation}\label{eq03}
\hat{H} = \frac{\hat{p}^2}{2m} + \frac{1}{2}m\omega_{\mathrm{eff}}^2\, \hat{x}^2 +
\frac{m\omega^2 a^2 \kappa}{2\pi^2},\qquad[\hat{x,\hat{p}}]=i\hbar
\end{equation}
We introduce the ladder operators,
\begin{equation}\label{eq04}
\hat{a} = \sqrt{\frac{m\omega_{\mathrm{eff}}}{2\hbar}}\,
\hat{x}\, + \,
\frac{i}{\sqrt{2m\hbar\omega_{\mathrm{eff}}}}\,\hat{p}, \qquad
\hat{a}^\dagger = \sqrt{\frac{m\omega_{\mathrm{eff}}}{2\hbar}}\,\hat{x}\,-\,
\frac{i}{\sqrt{2m\hbar\omega_{\mathrm{eff}}}}\,\hat{p},
\end{equation}
satisfying $[\hat{a},\hat{a}^\dagger]=1$. Substituting into Eq.~\ref{eq04} into Eq.~\ref{eq03}, with $\hat{N}=\hat{a}^\dagger\hat{a}$, the eigenvalue equation of the operator $\hat{H} \ket{n} =E_n \ket{n}$ yields:
\begin{equation}\label{eq05}
E_n = \hbar\,\omega\sqrt{1-\kappa}\left(n + \tfrac{1}{2}\right) +
\frac{m\omega^2 a^2 \kappa}{2\pi^2},
\quad n = 0, 1, 2, \dots
\end{equation}
The first term in  Eq.~\ref{eq05} is the quantized vibrational levels of the effective harmonic confinement, while the second term is a constant offset from the lattice contribution. This term shifts the quantized energy level;  however, maintains the system dynamics.  The corresponding normalized wavefunction: 
\begin{equation}\label{eq06}
    \psi_n(x) =\frac{1}{\sqrt{2^nn!}}\left(\frac{m\omega_{\text{eff}}}{\pi \hbar}\right)^{1/4} \mathcal{H}_n\left(\sqrt{\frac{m\omega_{\text{eff}}}{\hbar}}\right)\exp\left(-\frac{m\omega_{\text{eff}}}{\hbar} x^2\right)
\end{equation}
$\mathcal{H}_n$ is the Hermite polynomial of order $n$.
The corresponding wavefunction in momentum space using the Fourier transform \cite{robinett1995quantum}:
\begin{equation}\label{eq07}
    \psi_n(p)=\frac{1}{\sqrt{2 \pi \hbar}}\int \psi_n(x)\exp(-ipx/\hbar)\,dx
\end{equation}

The optical lattice modifies the system only through the effective frequency $\omega_{\mathrm{eff}}$, while the resulting eigenstates retain the harmonic-oscillator form. The distinction lies in how the effective frequency is controlled. 

\section{Fisher Information}\label{fisher}
Given the fixed harmonic structure established in Section~\ref{model}, 
Fisher information is introduced here as a diagnostic of wavefunction 
localization under effective frequency tuning. The general Fisher information for a spatial translation parameter is defined as the gradient functional of the probability density $\rho(x) = |\psi(x)|^2$ 
\cite{isonguyo2018quantum,luo2002fisher}:
\begin{equation}\label{eq08}
    I_x = \int \frac{[\partial_x \rho(x)]^2}{\rho(x)}\, dx 
        = 4\int |\partial_x \psi(x)|^2\, dx = 4\langle p^2\rangle,
\end{equation}
where the second equality follows because the eigenstates $\psi_n(x)$ 
in Eq.~\ref{eq06} are real-valued. For real wavefunctions, 
$\partial_x\rho = 2\psi\,\partial_x\psi$, so $I(\psi) = 4K(\psi)$ 
\cite{luo2002fisher}, where $K(\psi) = \int|\partial_x\psi|^2\,dx$ 
is the conventional kinetic energy \cite{romera2005fisher}. Eq.~\ref{eq08} measures the gradient content of the wavefunction and is directly proportional to the kinetic-energy density, which characterizes spatial localization 
\cite{luo2002fisher,stam1959some}. The variations in information 
measures arise solely from changes in the effective frequency 
$\omega_{\mathrm{eff}}$ through $\kappa$. Because effective frequency tightening enhances spatial localization, the momentum representation is included to quantify the complementary spreading. For $\rho(p) = |\psi(p)|^2$, the same real-valued reduction applies in momentum space, giving:
\begin{equation}\label{eq09}
    I_p = \int \frac{[\partial_p \rho(p)]^2}{\rho(p)}\, dp 
        = 4\int |\partial_p \psi(p)|^2\, dp = 4\langle x^2 \rangle.
\end{equation}
Using Eq.~\ref{eq06}, the Fisher information admits 
closed-form expressions in position and momentum space:
\begin{equation}\label{eq10}
   I_x = \frac{2m\omega\sqrt{1-\kappa}}{\hbar}(2n+1), \qquad 
I_p = \frac{2\hbar}{m\omega\sqrt{1-\kappa}}(2n+1).
\end{equation}
These expressions scale linearly with $\omega_{\mathrm{eff}}=\omega\sqrt{1-\kappa}$ 
and $(2n+1)$. This confirms that the effective frequency tuning 
redistributes localization between $x$ and $p$ conjugate spaces without 
changing the ladder structure. Their product,
\begin{equation}\label{fisherproduct}
    I_x I_p = 4(2n+1)^2,
\end{equation}
is independent of $\omega_{\mathrm{eff}}=\sqrt{1-\kappa}$ and satify the Cramér–Rao inequality~\cite{ dehesa2007fisher}. The uncertainty relation in Fig.~\ref{fisherproduct}, reveals that the  $I_x\,I_p$ product is bounded from below and increases with $n$.

We also consider the variance as a measure of the spreading of the probability density about its mean value. For an observable \( A\in\{x,p\} \), the variance is defined as
\begin{equation}\label{eq11}
\Delta A^2=\langle A^2\rangle-\langle A\rangle^2 .
\end{equation}

Using the standard harmonic-oscillator representation in terms of effective frequency $\omega_{\mathrm{eff}}$, the position and momentum variances for the eigenstates $|n\rangle$  is:
\begin{align}\label{eq12}
\Delta x^2 &= \frac{\hbar}{2\,m\,\omega\sqrt{1-\kappa}}\,(2n+1), \nonumber\\
\Delta p^2 &= \frac{m\,\hbar\,\omega\sqrt{1-\kappa}}{2}\,(2n+1).
\end{align}
These expressions show how effective frequency tuning through \( \kappa \) sharpens spatial localization while compensating momentum spreading. The product of the variances satisfies
\begin{equation}\label{eq13}
\Delta x\,\Delta p \ge \frac{\hbar}{2},
\end{equation}
which confirms that the effective frequency engineering redistributes localization between conjugate variables without altering the underlying harmonic structure.

\section{Fisher-Shannon complexity}\label{fishershannon}
We now consider the Fisher--Shannon complexity as a complementary measure that combines local gradient content with global spreading \cite{dehesa2007fisher}. The aim is to quantify how localization and delocalization balance as the effective frequency is varied. It is defined as:
\begin{align}\label{eq14}
   P_{\mathrm{FS}}^i =\mathcal{J}_i\, I_i, \\ \nonumber
   \mathcal{J}_{i} =\frac{1}{2 \pi e}\,e^{\frac{2}{D} S_i}
\end{align}
where $i\in\{x,p\}$, $D=1$, and $\mathcal{J}$ is the Shannon power~\cite{lopez2011statistical}. For a normalized probability density $\rho(x)$, the Shannon entropy in position and momentum space is defined as~\cite{shannon1948mathematical}
\begin{align}\label{eq15}
S_x = -\int \rho(x)\ln\,\rho(x)\,dx  \\ \nonumber
S_p = -\int \phi(p)\ln\,\phi(p)\,dp ,
\end{align}
where $\rho(x) = |\psi(x)|^2$ and $\phi(p)=|\psi(p)|$ are the probability densities in position and momentum spaces, respectively.
using the ground state wavefunction of Eqs.~\ref{eq06} and {}\ref{eq07}, the analytical expression for Shannon entropy in the ground state:
\begin{align}
   S_0(x) =\frac{1}{2}\,\left(1- \ln m\,-\ln \omega\sqrt{1-\kappa}\,+\, \ln \pi \hbar \right)\\ \nonumber
   S_0(p) = \frac{1}{2}\,\left(1\,+\,\ln \frac{m\,\pi\,\omega\sqrt{1-\kappa}}{\hbar}\right).
\end{align}

Because the effective frequency  $\omega_{\mathrm{eff}}=\sqrt{1-\kappa}$ can be tuned optically through $\kappa$, the Fisher-Shannon complexities $P_{\mathrm{FS}}^{(x,p)}$ both provide a symmetric map of localization balance under experimentally distinct control path.

\section{Discussion of Results}\label{result}

\subsection{Probability Density}
\begin{figure}[!ht]
    \centering
    
    \begin{minipage}[t]{0.48\linewidth}
        \centering
        \includegraphics[width=\linewidth]{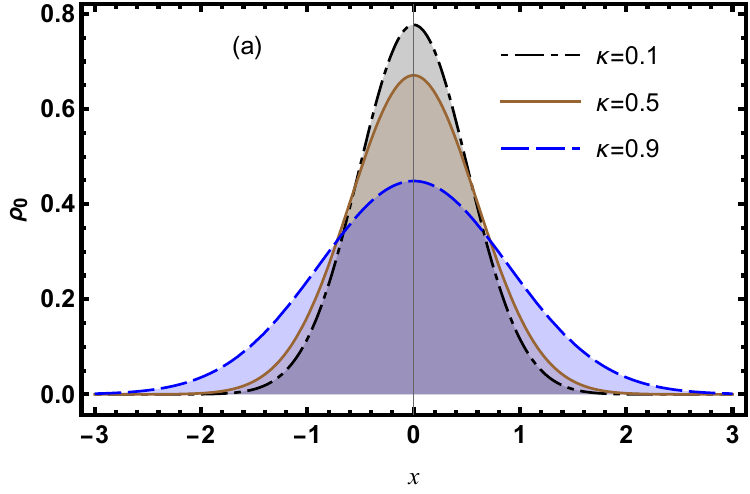}
    \end{minipage}
    \hfill
    \begin{minipage}[t]{0.48\linewidth}
        \centering
        \includegraphics[width=\linewidth]{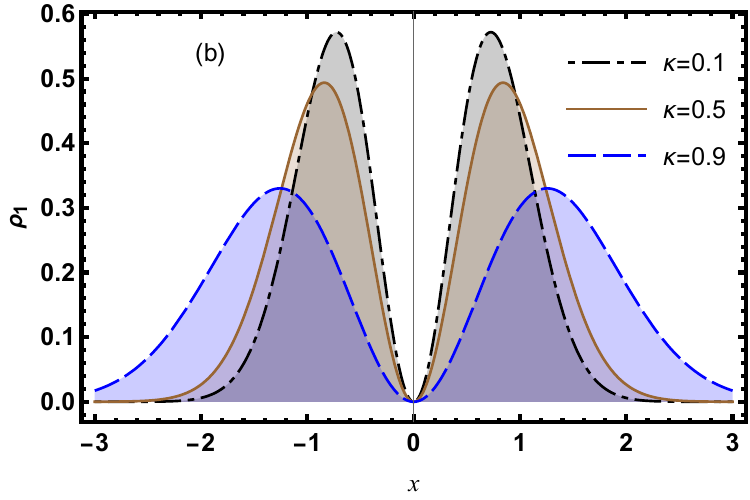}
    \end{minipage}
    \caption{ Probability density in position space $\rho(x)=|\psi(x)|^2$ for (a) ground state $n=0$ and (b) first excited state $(n=1)$, with $\omega=2$, $\kappa=0.1$ (dot-dashed), 
        $\kappa=0.5$ (solid), and $\kappa=0.9$ (dashed). $a=m=\hbar=1$}
    \label{Fig02}
\end{figure}

\begin{figure}[!ht]
    \centering
    
    \begin{minipage}[t]{0.48\linewidth}
        \centering
        \includegraphics[width=\linewidth]{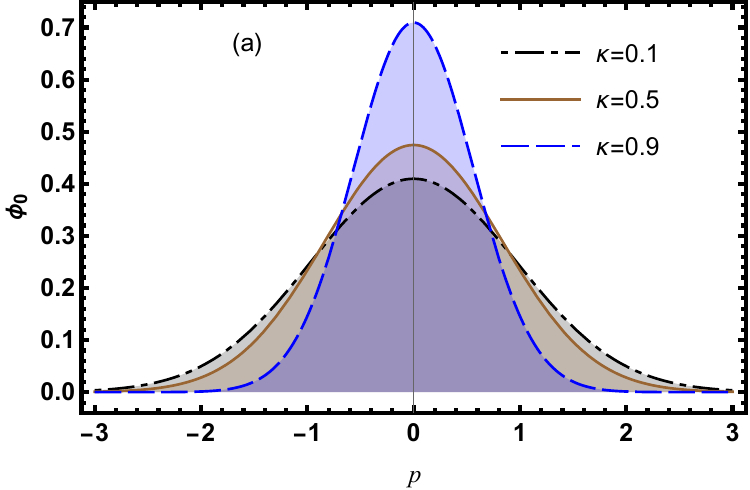}
    \end{minipage}
    \hfill
    \begin{minipage}[t]{0.48\linewidth}
        \centering
        \includegraphics[width=\linewidth]{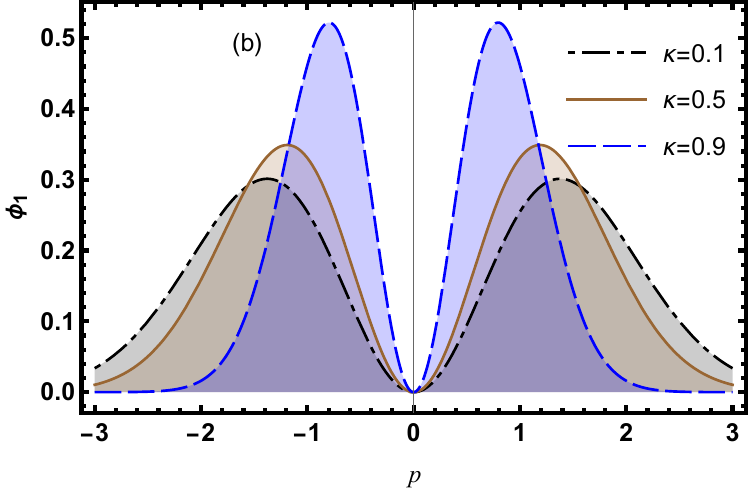}
    \end{minipage}
    \caption{Probability density in momentum space $\phi(p)=|\psi(p)|^2$ for (a) ground state and (b) first excited state. $\omega=2$, $\kappa=0.1$ (dot-dashed),  $\kappa=0.5$ (solid), and $\kappa=0.9$ (dashed) $a=m=\hbar=1$}
    \label{Fig03}
\end{figure}

In Fig.~\ref{Fig02} and {}~\ref{Fig03}, we plot the probability density in position and momentum spaces, respectively. We observe how varying the parameter $\kappa$ at constant $\omega=2$ affects the probability distribution. In position space, Fig.~\ref{Fig02}, increasing $\kappa$ broadens $\rho_0$ and $\rho_1$ while lowering their maxima, but the response differs by state. The ground state spreads smoothly, whereas the excited state shows enhanced separation of its lobes and a sharper node. This contrast confirms that effective frequency softening acts globally; however, the nodal structure amplifies its effect in excited states, where effective frequency tuning competes with nodal constraints rather than Gaussian widening. In momentum space, Fig.~\ref{Fig03}, increasing $\kappa$ sharpens $\phi_0$ and $\phi_1$ and raises their peak heights. Again, the excited state reacts more strongly. The ground state tightens uniformly, while the excited state develops narrower, higher peaks separated by a deeper minimum, which reflect stronger momentum localization tied to nodal structure. The Optical tuning through $\kappa$ alters the effective frequency without breaking harmonicity; but it reshapes spatial and momentum localization continuously and state-dependently. This effective frequency-controlled redistribution has been observed in trapped-ion lattices where optical potentials modify wavefunction support without changing ladder structure~\cite{gangloff2015velocity}. Our result establishes $\kappa$ as the dominant experimental handle for localization control in the Paul-trap lattice framework. Varying $\kappa$ tunes the effective frequency, shifts localization between conjugate spaces, and affects excited states more strongly. $\kappa$ reshapes how the nodes and effective frequency interact.


\subsection{Fisher Information and Variance}\label{sec:fisher}

\begin{table}[!t]
\centering
\caption{Fisher information, expectation value and variance for $\omega =1$}
\begin{tabular}{|c|c|c|c|c|c|c|c|}\hline
\hline
$n$ & $\kappa$ & $I_{x}$ & $I_{p}$ & $I_xI_p$ & $\braket{x^2}$ & $\braket{p^2}$ & $\Delta x \Delta p$\\ \hline
\hline
  & 0.2  & 1.78885   & 2.23607  & 4.0000 & 0.559017 & 0.447214 & 0.5000 \\
0 & 0.4  & 1.54919   & 2.58199  & 4.0000 & 0.645497 & 0.387298 & 0.5000 \\
  & 0.8  & 0.894427  & 4.47214  & 4.0000 & 1.11803  & 0.223607 & 0.5000 \\ 
  \hline
  & 0.2  & 5.36656   & 6.7082   & 36.0000 & 1.67705 & 1.34164 & 1.5000 \\
1 & 0.4  & 4.64758   & 7.74597  & 36.0000 & 1.93649 & 1.1619  & 1.5000 \\
  & 0.8  & 2.68328   & 13.4164  & 36.0000 & 3.3541  & 0.67082 & 1.5000 \\
  \hline
  & 0.2  & 8.94427  & 11.1803  & 100.0000 & 2.79508 & 2.23607 & 2.5000 \\
2 & 0.4  & 7.74597  & 12.9099  & 100.0000 & 3.22749 & 1.93649 & 2.5000 \\
  & 0.8  & 4.47214  & 22.3607  & 100.0000 & 5.59017 & 1.11803  & 2.5000 \\
  \hline
  & 0.2  & 8.94427  & 15.6525  & 196.0000 & 3.91312 & 3.1305  & 3.5000 \\
3 & 0.4  & 7.74597  & 18.0739  & 196.0000 & 4.51848 & 2.71109 & 3.5000 \\
  & 0.8  & 4.47214  & 31.305   & 196.0000 & 7.82624 & 1.56525 & 3.5000 \\ 
  \hline
\end{tabular}\label{tab1}
\end{table}

In Table~\ref{tab1}, we present the numerical results for Fisher information, expectation value and variance under the tunning $\kappa$ parameter. We observe that for every $n$, increasing $\kappa$, reduces the effective frequency $\omega_{\mathrm{eff}}$, which leads delocalization of the position-space Fisher information $I_x$ and localization of the momentum space Fisher information $I_p$. Softening the trap compresses the momentum distribution, which increases its gradients and raises $I_p$. Every row of the table reflects this inversion: entries of $I_p$ grow as $I_x$ shrink.

The numerical result also show that for every fixed eigenstate $n$, the product of the position and momentum-space Fisher information is invariant under the $\kappa$ parameter sweep. The numerical values in Table~\ref{tab1} are consistent with these analytical expressions of Eq.~\ref{fisherproduct} for all $n$ and $\kappa$ values reported.

Furthermore, the table shows that the expectation values $\langle x^{2}\rangle$ increases and  $\langle p^{2}\rangle$ decreases for increasing $\kappa$( and the trap effective frequency decreases). However, both cases scale linearly with $2n+1$, which is expected for a harmonic spectrum. Also, their monotonicity ensures that the uncertainty product remains $\Delta x\,\Delta p = n\,+\, 1/2$ for all parameters. This demonstrates that $\kappa$ sweep introduces isotropic scaling of the oscillator phase space. Therefore, all observables comprising Fisher information, variances, and uncertainty, follow this single parameter, which gives a consistent diagnostic of how effective frequency engineering reshapes localization in the trapped-ion system.

\subsection{Fisher-Shannon Complexity}\label{complexity}

\begin{figure}[!t]
    \centering

    \begin{minipage}[t]{0.48\linewidth}
        \centering
        \includegraphics[width=\linewidth]{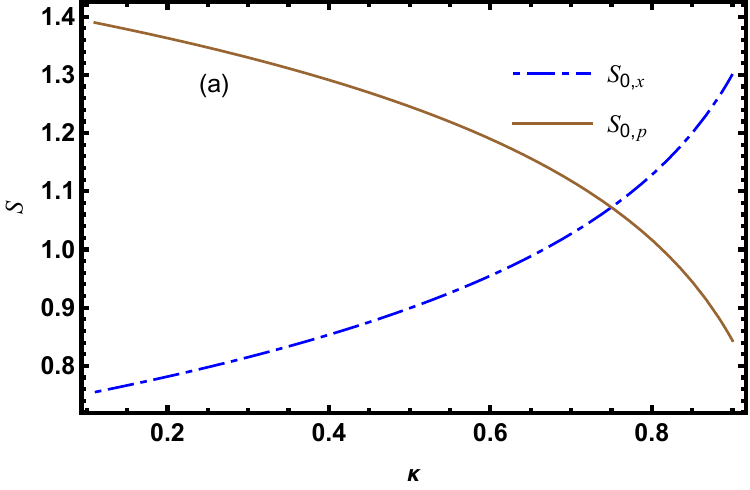}
    \end{minipage}
    \hfill
    \begin{minipage}[t]{0.48\linewidth}
        \centering
        \includegraphics[width=\linewidth]{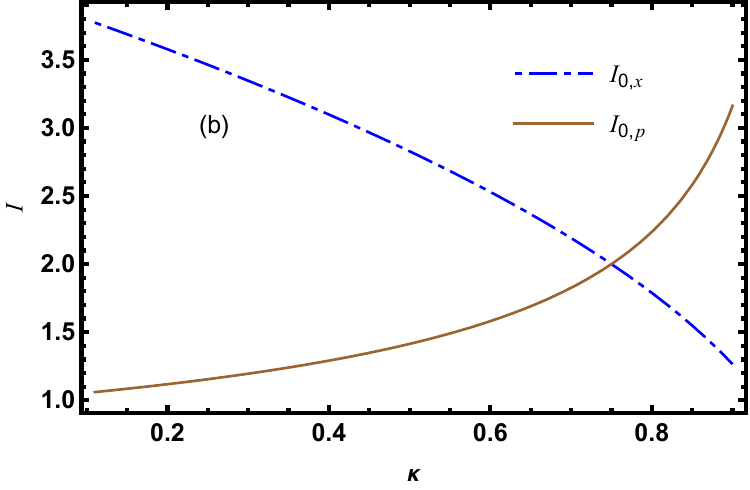}
    \end{minipage}

    \caption{Shannon entropy (a) and Fisher information (b) in position space (blue dot-dashed) and momentum space (brown solid) as functions of the $\kappa$ in the ground state. Increasing $\kappa$ reduces $\omega_{\mathrm{eff}}$ at constant $\omega=2$, which drives $S_{0,x}$ up and $S_{0,p}$ down in (a), and suppresses $I_{0,x}$ while raising $I_{0,p}$ in (b), redistributing information between conjugate spaces.}
    \label{fig04}
\end{figure}

Figure~\ref{fig04}(a) shows that the position- and momentum-space Shannon entropies respond oppositely to the effective frequency $\omega_\mathrm{eff}$ through $\kappa$. As $\omega_\mathrm{eff}$ decreases with increasing $\kappa$, the position-space $S_{0,x}$ entropy increasing monotonically. This reflects a weakened spatial confinement and increased configurational uncertainty. At the same time, the momentum-space entropy $S_{0,p}$  decreases, which broadens the momentum distribution and increases localization. These trends do not represent independent effects; they are the direct consequence of harmonic rescaling controlled by a single parameter $\kappa$. Effective frequency, therefore, reallocates uncertainty between conjugate variables while preserving the global entropic balance required by quantum uncertainty relations.

Figure~\ref{fig04}(b) shows a complementary pattern for the Fisher information. From Eq.~\ref{eq10}, $I_{0,x} = (2m\omega_{\mathrm{eff}}/\hbar)(2n+1)$ scales linearly with $\omega_{\mathrm{eff}}$. For decreasing  $\omega_{\mathrm{eff}}$, the position-space Fisher information decreases, reflecting particle delocalization. In contrast, $I_{0,p} = (2\hbar/m\omega_{\mathrm{eff}})(2n+1)$ scales inversely with 
$\omega_{\mathrm{eff}}$, so momentum-space Fisher information increases monotonically, signaling particle localization. These opposing trends arise directly from the harmonic rescaling in Eq.~\ref{eq10} and are fully consistent with Fig.~\ref{fig04}(B). Fisher and Shannon measures, therefore, respond to effective frequency as consistent but partial indicators, encoding how information is redistributed locally and globally without introducing new structural content. Since $\omega_{\mathrm{eff}} = \omega\sqrt{1-\kappa}$, increasing $\kappa$ softens the trap and drives $I_x$ down while raising $I_p$.

\begin{figure}[!t]
    \centering

    \begin{minipage}[t]{0.48\linewidth}
        \centering
        \includegraphics[width=\linewidth]{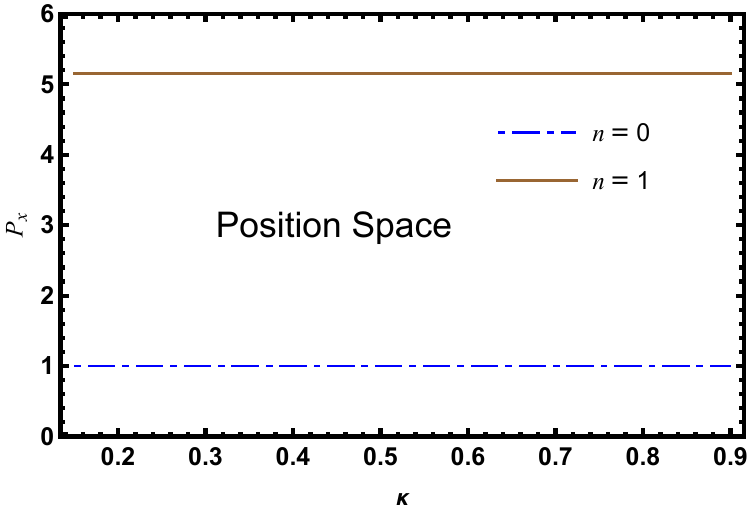}
    \end{minipage}
    \hfill
    \begin{minipage}[t]{0.48\linewidth}
        \centering
        \includegraphics[width=\linewidth]{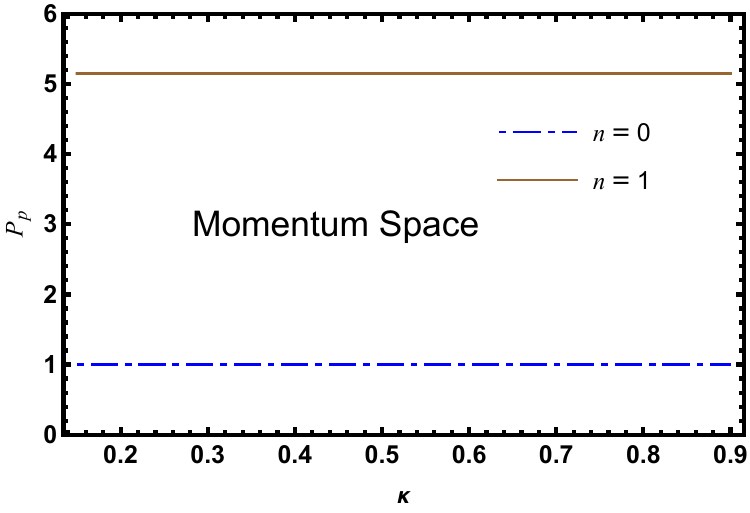}
    \end{minipage}

    \caption{Fisher-Shannon complexity measures in (a) position space and (b) momentum space as function of $\kappa$, for $n=0$ (blue dot dashed) and $n=1$ (brown solid). Even with the opposing trends in Fig.~\ref{fig04}, the complexity remains strictly invariant under $ \omega_{\mathrm{eff}} = \omega\sqrt{1-\kappa}$ tuning, which confirms that optical lattice control $\kappa$ rescales localization without altering the structural content of the motional states.}
    \label{fig05}
\end{figure}

A qualitatively different outcome emerges for the Fisher–Shannon complexity measure, shown in Fig.~\ref{fig05}. Despite the clear effective frequency dependence through $\kappa$, Fisher–Shannon complexity measure remains invariant in both position and momentum space. For the ground state, we find the minimal value $P_x = P_p=1$, independent of $\omega_\mathrm{eff}$ or, equivalently, of the lattice control parameter $\kappa$. Importantly, the same flat behavior persists for the first excited state. We conclude that the excitation changes the absolute level of complexity, but not its response to harmonic effective frequency modulation.

The invariance of $P$ has a clear physical interpretation: effective frequency control alone reshapes scale but not structure. While increasing $\omega_\mathrm{eff}$ sharpens position and broadens momentum, the nodal topology and internal correlations of harmonic eigenstates remain unchanged, and the Fisher–Shannon complexity reflects this rigidity. Any deviation from the flat behavior reported here would therefore constitute unambiguous evidence of physics beyond effective harmonic confinement, such as genuine lattice-induced anharmonicity, mode mixing, or non-pure motional states. In this sense, the observed invariance provides a stringent reference against which more complex trapped-ion and lattice-driven regimes can be assessed.
\subsection{Breakdown of Invariance Beyond the Harmonic limit}

The harmonic analysis in Sec.~\ref{sec:fisher} and~\ref{complexity} is grounded in the small-oscillation approximation, where the ion remains tightly localized near $x=0$ and the lattice potential is well described by its quadratic component alone. Extending this description to larger oscillation amplitudes, we introduce higher order term from Eq.~\ref{Eq01}. Here, the ion begins to sample the anharmonic curvature of the lattice. The effective anharmonic potential becomes

\begin{equation}\label{eq18}
   V_{\mathrm{anh}}(x) = \frac{m\omega^2\pi^2\kappa}{6a}x^4 
    + \frac{m\omega^2}{2}(1-\kappa)x^2 
    + \frac{m\omega^2 a^2\kappa}{2\pi^2} + O(x^6),
\end{equation}
where, the constant offset shifts all energy levels uniformly. The quadratic term preserves the Gaussian structure of the harmonic eigenstates and the quartic correct term breaks it~\cite{dong2019exact}.
We therefore treat the quartic term as a perturbation~\cite{bender1973nharmonic} (See Appendix~\ref{appendix:perturbation} for detailed derivation), the first-order corrected wavefunctions for the ground and first excited states are:
\begin{align}\label{eq19}
\psi_0(x) &\approx \mathcal{N}_0\Bigl[
\psi_0^{(0)}(x) 
- 3\sqrt{2}\,\lambda\,\psi_2^{(0)}(x)
- \frac{\sqrt{6}}{2}\,\lambda\,\psi_4^{(0)}(x) \Bigr], 
\nonumber\\
\psi_1(x) &\approx \mathcal{N}_1\Bigl[
\psi_1^{(0)}(x) 
- 5\sqrt{6}\,\lambda\,\psi_3^{(0)}(x)
- \frac{\sqrt{30}}{2}\,\lambda\,\psi_5^{(0)}(x)\Bigr],
\end{align}
where $\lambda$ and $\mathcal{N}_{0,1}$ are defined in~\ref{eq:lambda}, {}\ref{eq:psi0_corrected}, and {}\ref{eq:psi1_corrected} respectively. $\psi_{0,1}^{(0)}$ are the unperturbed harmonic wavefunctions. The anharmonic wavefunctions are therefore non-Gaussian, carrying oscillatory structure and heavier tails in magnitude.  

\begin{figure}[ht!]
    \centering

    \begin{minipage}[t]{0.48\linewidth}
        \centering
        \includegraphics[width=\linewidth]{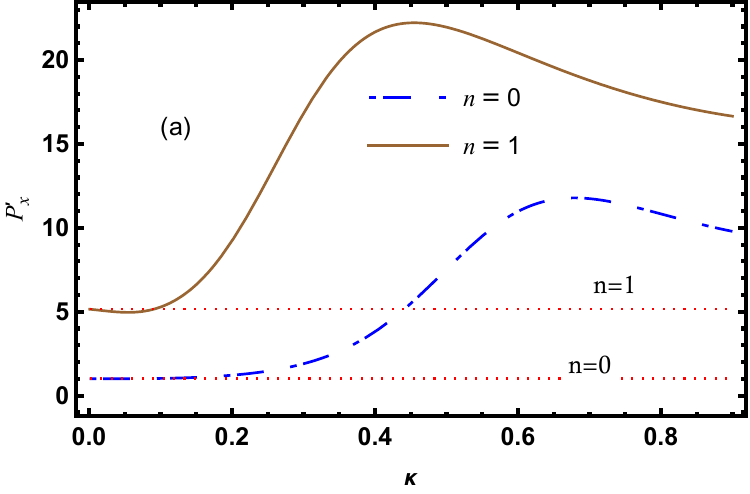}
    \end{minipage}
    \hfill
    \begin{minipage}[t]{0.48\linewidth}
        \centering
        \includegraphics[width=\linewidth]{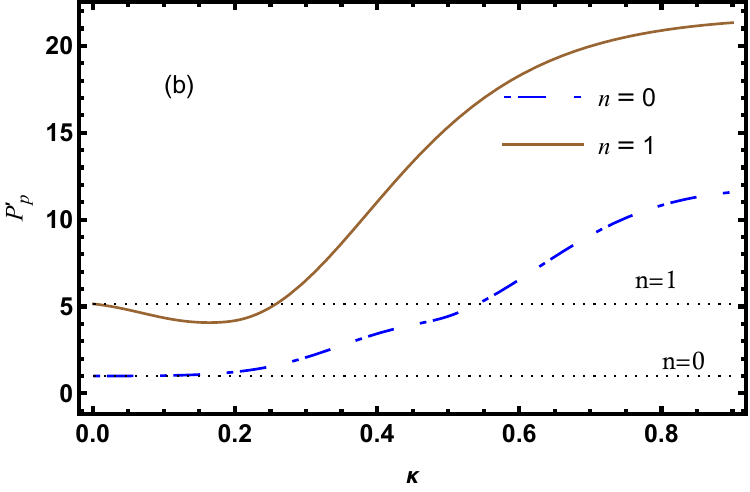}
    \end{minipage}
    \caption{Fisher-Shannon complexity $P^\prime$ in (a) position space and (b) momentum space, computed using the first-order perturbation-corrected 
wavefunctions in Eq.~\ref{eq19} for $n=0$ (blue dot-dashed) and $n=1$ (brown solid), with $m=\hbar=a=1$ and $\omega=2$. Dotted horizontal lines mark the harmonic invariant values from Sec~\ref{complexity}. Both states depart from their harmonic reference as $\kappa$ increases, with $n=1$ departing earlier and more strongly, confirming that the quartic correction breaks the complexity 
invariance in a state-dependent manner.}
    \label{fig06}
\end{figure}

The invariance established in Sec.~\ref{complexity} rests on the mutual compensation of Fisher information and Shannon entropy under harmonic rescaling. Fisher information tracks local gradient structure; Shannon entropy tracks global spreading. In the harmonic case, Fisher information and Shannon entropy scale in a mutually compensating manner, sustaining the complexity invariant. The quartic correction breaks this compensation, so $P$ is no longer invariant. Figure~\ref{fig06} confirms this. In position space, both states depart upward from their harmonic reference as $\kappa$ increases, with $n=1$ departing earlier and reaching larger values. In momentum space, both states show an initial dip followed by monotonic growth, again with $n=1$ more sensitive. The state-dependence of the departure is consistent with the larger mixing coefficients of $\psi_1$: $c_3 = -5\sqrt{6}\,\lambda$ exceeds $c_2 = -3\sqrt{2}\,\lambda$ in magnitude at the same $\lambda$, which produces a stronger structural deformation of the excited state. The onset of deviation from the flat reference marks the boundary at which the strong-confinement approximation breaks down.

The invariance established here serves as a theoretical reference, not a direct experimental protocol. In practice, deviations from the predicted invariant values could arise from: anharmonicity and trap imperfection, imperfect motional state preparation, or measurement inaccuracy in reconstructing the probability density~~\cite{mccormick2019quantum, lin2020quantum,nguyen2010sums}. Distinguishing these sources would require independent characterization of the motional state purity, for instance via sideband thermometry or Wigner function tomography \cite{leibfried2003quantum}. The present results, therefore, provide a theoretical boundary condition: within the strong-confinement harmonic regime, the invariants hold exactly, and any departure that survives state characterization can be attributed to non-harmonic lattice physics. This is complementary to, not competitive with, direct spectroscopic probing of anharmonic frequency shifts.


\section{Conclusion}\label{conclusion}
We analyze the informational structure of a trapped ion governed by the effective frequency of the effective potential, $\omega_{\mathrm{eff}}=\omega\, \sqrt{1-\kappa}$. Our results show that Fisher information and Shannon entropy respond coherently and predictably to changes in confinement strength. Softening the trap reduces position-space gradients, suppressing Fisher information while increasing Shannon entropy, with complementary behavior in momentum space. These opposing trends confirm that effective frequency modulation redistributes information between conjugate variables without altering the underlying harmonic structure of the oscillator. While the individual information-theoretic results follow from known properties of harmonic eigenstates \cite{sakurai2017modern}, their realization under optical $\kappa$ control in a lattice-assisted Paul trap is new, and the invariants provide a reference against which genuinely non-harmonic lattice effects can be diagnosed.

The optical lattice, therefore, acts as an effective frequency-control mechanism rather than a source of new information content. This invariance establishes a clear baseline: any deviation from it would signal genuinely non-harmonic effects or higher-order lattice contributions. As such, the present results provide a robust reference framework for assessing more complex lattice-assisted Paul trap configurations.

\section*{Funding}
This research is support from the NSRF via the Program Management Unit for Human Resources \& Institutional Development, Research and Innovation [grant number B39G680007]

\section*{Acknowledgment}
This research was supported by the Ph.D Student Exchange Scholarship of Prince of Songkla University

\section*{Data Availability}
All data generated or analysed during this study are included in this published article

\section*{Declaration of competing interest}
The authors declare no competing interests.
\section*{Appendix}
\appendix
\section{Perturbation Theory}
\label{appendix:perturbation}

\subsection{Hamiltonian decomposition}

The full lattice potential in Eq.~\ref{Eq01}, expanded to fourth order 
in $x$, gives the Hamiltonian

\begin{equation}
\hat{H} = \hat{H}_0 + \hat{H}' + C,
\end{equation}

where the unperturbed harmonic part is

\begin{equation}
\hat{H}_0 = \frac{\hat{p}^2}{2m} + \frac{1}{2}m\omega_{\mathrm{eff}}^2\hat{x}^2,
\qquad \omega_{\mathrm{eff}} = \omega\sqrt{1-\kappa},
\end{equation}

the quartic perturbation is:

\begin{equation}
\hat{H}' = \beta\hat{x}^4, \qquad 
\beta = \frac{m\omega^2\pi^2\kappa}{6a},
\end{equation}

and $C = m\omega^2 a^2\kappa/2\pi^2$ is a constant offset that shifts 
all energy levels uniformly without affecting eigenstates. It is retained in the energy expression 
but dropped throughout the wavefunction analysis.

The unperturbed eigenstates $|n\rangle$ satisfy 
$\hat{H}_0|n\rangle = E_n^{(0)}|n\rangle$ with

\begin{equation}
E_n^{(0)} = \hbar\omega_{\mathrm{eff}}\left(n+\frac{1}{2}\right).
\end{equation}

We introduce the ladder operators

\begin{equation}
\hat{a} = \sqrt{\frac{m\omega_{\mathrm{eff}}}{2\hbar}}\,\hat{x} 
          + \frac{i}{\sqrt{2m\hbar\omega_{\mathrm{eff}}}}\,\hat{p}, 
\qquad
\hat{a}^\dagger = \sqrt{\frac{m\omega_{\mathrm{eff}}}{2\hbar}}\,\hat{x} 
                 - \frac{i}{\sqrt{2m\hbar\omega_{\mathrm{eff}}}}\,\hat{p},
\end{equation}

satisfying $[\hat{a},\hat{a}^\dagger]=1$, so that $\hat{x} = \alpha\,
(\hat{a}+\hat{a}^\dagger)$, 
where $\alpha=\sqrt{\hbar/2m\omega_{\mathrm{eff}}}$. The dimensionless perturbation strength is defined as

\begin{equation}\label{eq:lambda}
\lambda = \frac{\beta\hbar}{4m^2\omega_{\mathrm{eff}}^3} 
        = \frac{\pi^2\kappa\hbar\omega^2}{24am\omega_{\mathrm{eff}}^3}
        = \frac{\pi^2\kappa\hbar}{24am\omega(1-\kappa)^{3/2}}
\end{equation}

\subsection{Ground state (n=0)}
We consider the action of $\hat{x}^4$ on the ground state ($n=0$):
\begin{equation}
   (\hat{a}+\hat{a}^\dagger)^4\ket{0} =3|0\rangle + 6\sqrt{2}|2\rangle + 2\sqrt{6}|4\rangle
\end{equation}
The off-diagonal matrix elements of $\hat{H}'$ needed for perturbation theory are therefore

\begin{align}
\langle 2|\hat{H}'|0\rangle &= 6\sqrt{2}\,\beta\alpha^2,\label{eq:H2}\\ \nonumber
\langle 4|\hat{H}'|0\rangle &= 2\sqrt{6}\,\beta\alpha^2.\label{eq:H4}
\end{align}

The first-order correction to the ground state energy is

\begin{equation}
E_0^{(1)} = \langle 0|\hat{H}'|0\rangle 
           = 3\beta\alpha^2 
           = \frac{3\beta\hbar^2}{4m^2\omega_{\mathrm{eff}}^2} = \frac{\pi^2\kappa\hbar^2}{8am(1-\kappa)}.
\end{equation}


The second-order correction receives contributions only from 
$|2\rangle$ and $|4\rangle$, since those are the only states 
connected to $|0\rangle$ through $\hat{H}'$:

\begin{align}
E_0^{(2)} 
&= \frac{|\langle 2|\hat{H}'|0\rangle|^2}{E_0^{(0)}-E_2^{(0)}} 
 + \frac{|\langle 4|\hat{H}'|0\rangle|^2}{E_0^{(0)}-E_4^{(0)}} 
\nonumber\\
&= \frac{(6\sqrt{2}\,\beta\alpha^2)^2}{-2\hbar\omega_{\mathrm{eff}}} 
 + \frac{(2\sqrt{6}\,\beta\alpha^2)^2}{-4\hbar\omega_{\mathrm{eff}}} 
= -\frac{21\beta^2\hbar^3}{8m^4\omega_{\mathrm{eff}}^5}
\end{align}

The total ground state energy, including the lattice constant 
offset $C$, is

\begin{equation}\label{eq:E0_full}
E_0 = \frac{\hbar\omega_{\mathrm{eff}}}{2} 
    + \frac{m\omega^2 a^2\kappa}{2\pi^2}
    + \frac{3\beta\hbar^2}{4m^2\omega_{\mathrm{eff}}^2}
    - \frac{21\beta^2\hbar^3}{8m^4\omega_{\mathrm{eff}}^5}
    + O(\lambda^3).
\end{equation}

The first term is the harmonic zero-point energy under effective frequency $\omega_{\mathrm{eff}}$. The second is the lattice offset. The third is the positive first-order anharmonic shift, and the fourth is the negative second-order correction, which grows as $\kappa^2$ for small $\kappa$.


Next, we solve for the first-order correction to the ground state wavefunction defined as

\begin{equation}
|\psi_0^{(1)}\rangle = \sum_{n\neq 0}
\frac{\langle n|\hat{H}'|0\rangle}{E_0^{(0)}-E_n^{(0)}}|n\rangle
= c_2|2\rangle + c_4|4\rangle,
\end{equation}

with mixing coefficients

\begin{align}
c_2 &= \frac{6\sqrt{2}\,\beta\alpha^2}{-2\hbar\omega_{\mathrm{eff}}} 
     = -3\sqrt{2}\,\lambda,\\ \nonumber
c_4 &= \frac{2\sqrt{6}\,\beta\alpha^2}{-4\hbar\omega_{\mathrm{eff}}} 
     = -\frac{\sqrt{6}}{2}\,\lambda.
\end{align}

The corrected ground state wavefunction, normalised to first order 
in $\lambda$, is

\begin{equation}\label{eq:psi0_corrected}
\psi_0(x) \approx \mathcal{N}_0\Bigl[
\psi_0^{(0)}(x) - 3\sqrt{2}\,\lambda\,\psi_2^{(0)}(x)
- \frac{\sqrt{6}}{2}\,\lambda\,\psi_4^{(0)}(x)\Bigr],
\end{equation}
where $\mathcal{N}_0 = (1 + 18\lambda^2 + 3\lambda^2/2)^{-1/2}$ and $\psi_n^{(0)}(x)$ is the unperturbed harmonic eigenfunctions.



The corrected wavefunction in Eq.~\ref{eq:psi0_corrected} is no 
longer a pure Gaussian. Both coefficients $c_2$ and $c_4$ grow with $\kappa$ through $\lambda$, so the non-Gaussian character of $\psi_0$ strengthens as the optical lattice depth increases.

\subsection{First excited state \texorpdfstring{$n=1$}{ex}}
We consider the action of $\hat{x}^4$ on the first excited state ($n=1$):
\begin{equation}
   (\hat{a}+\hat{a}^\dagger)^4\ket{1} =15|1\rangle + 10\sqrt{6}|3\rangle + 2\sqrt{30}|5\rangle
\end{equation}

The perturbation $\hat{H}' = \beta\hat{x}^4 = \beta\alpha^2
(\hat{a}+\hat{a}^\dagger)^4$ therefore connects $|1\rangle$ only 
to $|3\rangle$ and $|5\rangle$. The nonzero off-diagonal matrix 
elements are:

\begin{align}
\langle 3|\hat{H}'|1\rangle &= 10\sqrt{6}\,\beta\alpha^2, 
\label{eq:H31}\\
\langle 5|\hat{H}'|1\rangle &= 2\sqrt{30}\,\beta\alpha^2.
\label{eq:H51}
\end{align}

To solve for the first-order energy correction, the diagonal matrix element gives the first-order shift:

\begin{equation}
E_1^{(1)} = \langle 1|\hat{H}'|1\rangle 
           = 15\beta\alpha^2 
           = \frac{15\beta\hbar^2}{4m^2\omega_{\mathrm{eff}}^2} = \frac{5\pi^2\kappa\hbar^2}{8am(1-\kappa)}.
\end{equation}

We can see that $E_1^{(1)} = 5E_0^{(1)}$, which reflects the stronger 
sensitivity of the excited state to anharmonic confinement.

We solve for second-order energy correction. Second-order contributions come only from the states $|3\rangle$ and 
$|5\rangle$:

\begin{align}
E_1^{(2)} 
&= \frac{|\langle 3|\hat{H}'|1\rangle|^2}{E_1^{(0)}-E_3^{(0)}} 
 + \frac{|\langle 5|\hat{H}'|1\rangle|^2}{E_1^{(0)}-E_5^{(0)}} 
\nonumber\\
&= \frac{(10\sqrt{6}\,\beta\alpha^2)^2}{-2\hbar\omega_{\mathrm{eff}}} 
 + \frac{(2\sqrt{30}\,\beta\alpha^2)^2}{-4\hbar\omega_{\mathrm{eff}}} = -\frac{330\beta^2\hbar^3}{16m^4\omega_{\mathrm{eff}}^5}.
\end{align}

The first excited state energy to second order is given as
\begin{equation}\label{eq:E1_full}
E_1 = \frac{3\hbar\omega_{\mathrm{eff}}}{2} 
    + \frac{m\omega^2 a^2\kappa}{2\pi^2}
    + \frac{15\beta\hbar^2}{4m^2\omega_{\mathrm{eff}}^2}
    - \frac{330\beta^2\hbar^3}{16m^4\omega_{\mathrm{eff}}^5}
    + O(\lambda^3).
\end{equation}

Comparing with Eq.~\ref{eq:E0_full}, the anharmonic shift scales 
as $15\beta\alpha^2$ for $n=1$ and $3\beta\alpha^2$ for $n=0$. 
The ratio of five confirms that the excited state is five times 
more sensitive to the quartic correction at first order.

We solve for the first-order wavefunction correction.The mixing coefficients are:

\begin{align}
c_3 &= \frac{\langle 3|\hat{H}'|1\rangle}
             {E_1^{(0)}-E_3^{(0)}} 
     = \frac{10\sqrt{6}\,\beta\alpha^2}{-2\hbar\omega_{\mathrm{eff}}} 
     = -5\sqrt{6}\,\lambda, \label{eq:c3}\\
c_5 &= \frac{\langle 5|\hat{H}'|1\rangle}
             {E_1^{(0)}-E_5^{(0)}} 
     = \frac{2\sqrt{30}\,\beta\alpha^2}{-4\hbar\omega_{\mathrm{eff}}} 
     = -\frac{\sqrt{30}}{2}\,\lambda. \label{eq:c5}
\end{align}

The corrected first excited state, normalized to first order in 
$\lambda$, is:

\begin{equation}\label{eq:psi1_corrected}
\psi_1(x) \approx \mathcal{N}_1\Bigl[
\psi_1^{(0)}(x) 
- 5\sqrt{6}\,\lambda\,\psi_3^{(0)}(x)
- \frac{\sqrt{30}}{2}\,\lambda\,\psi_5^{(0)}(x)
\Bigr],
\end{equation}

where the normalisation constant to second order in $\lambda$ is:

\begin{equation}
\mathcal{N}_1 = \frac{1}{\sqrt{1 + 150\lambda^2 
+ \frac{15}{2}\lambda^2}}
= \frac{1}{\sqrt{1 + \frac{315}{2}\lambda^2}}.
\end{equation}

Comparing Eq.~\ref{eq:psi1_corrected} with Eq.~\ref{eq:psi0_corrected}, the parity structure is preserved: odd unperturbed states mix only with odd higher states, and even with even. The coefficients $c_3 = -5\sqrt{6}\,\lambda$ and $c_5 = -\sqrt{30}\lambda/2$ are both larger in magnitude than $c_2=3\,\sqrt{2}\,\lambda$ and $c_4=\sqrt{6}\,\lambda/2$ of the ground state at the same $\lambda$, which confirms stronger structural deformation of the excited state under anharmonic correction.
\bibliographystyle{elsarticle-num} 
\bibliography{reference}
\end{document}